\documentclass[11pt,a4paper]{article}

\usepackage{amsmath}
\usepackage{amsthm,amssymb}

\usepackage{graphicx} 

\newcommand{\beq}{\begin{equation}}
\newcommand{\eeq}{\end{equation}}

\numberwithin{equation}{section}

\newcommand{\qbinom}[3]{\genfrac{[}{]}{0pt}{}{#1}{#2}_{#3}}

\begin{document}

\title{A simple model of a vesicle drop in a confined geometry}
\author{A.L.~Owczarek$^1$ and T. Prellberg$^2$ \\[1ex]
\footnotesize
  \begin{minipage}{13cm} 
 $^1$Department of Mathematics and Statistics,\\
  The University of Melbourne,\\Parkville, Victoria 3010, Australia\\[1ex]
  $^2$ School of Mathematical Sciences\\
Queen Mary University of London\\
Mile End Road, London E1 4NS, UK
\end{minipage}
}

\maketitle

\begin{abstract}

  We present the exact solution of a two-dimensional directed walk
  model of a drop, or half vesicle, confined between two walls, and
  attached to one wall. This model is also a generalisation of a polymer model
  of steric stabilisation recently investigated.  We explore the
  competition between a sticky potential on the two walls and the
  effect of a pressure-like term in the system. We show that a
  negative pressure ensures the drop/polymer is unaffected by
  confinement when the walls are a macroscopic distance apart. 
\end{abstract}

\section{Introduction}

The study of the behaviour of the boundary between two phases has a
long history \cite{temperley1952a-a,dietrich1988a-a}. In
two-dimensions where the boundary is one-dimensional much work has
been done to understand such behaviour \cite{fisher1984a-a}. One type
of model that has proved useful involves directed walks in a half
plane with a Boltzmann weight associated with the area under the walk
\cite{owczarek1993a-:a}.  Related to this is the study of an enclosed
boundary that can be used to model biological membranes \cite{fisher1991a-a}. The study of
lattice vesicles is motivated by biological membranes that
consist of lipid bi-layers and form shapes that depend on acidity, osmotic
pressure and temperature.  The basic model in two dimensions consists
of some type of self-avoiding polygon on a lattice which is weighted
according to its area and perimeter. The weighting of the area is
analogous to an osmotic pressure. Various exactly solved cases
including directed walk model have been considered \cite{brak1990a-a,brak1994a-:a,prellberg1995c-:a}.

In a parallel development the study of the behaviour of a long linear
polymer molecule in dilute solution confined between two parallel
plates \cite{janse2005a-:a} has recently gained momentum with the
exact solution of various directed walk models
\cite{brak2005a-:a,owczarek2008c-:a}. The phenomena being modelled
here are the steric stabilisation and sensitised flocculation of
colloidal dispersions. The progress made by the exact solution is the
finding that if one considers a polymer confined between two sticky
walls, the thermodynamic limit where the polymer is much longer than
the distance between the walls is shown to have a different phase
structure to the case where is there is one wall only \cite{brak2005a-:a}.

In this paper we study a directed version of the self-avoiding vesicle
model that is a generalisation of both the vesicle models 
studied previously \cite{prellberg1995c-:a} and the recently studied
polymer model \cite{brak2005a-:a}.  It is effectively a
half-vesicle, or drop, if considered as a phase boundary. We consider
directed self-avoiding walks on the square lattice, confined between
two lines ($y=0$ and $y=w$) whose ends are both attached to one of the
lines so that it effectively forms a polygon or rather a \emph{loop}.
We add weights for visits of the walk to both the top wall and the
bottom wall. We also add an (osmotic) pressure-like term that weights the
area contained under the loop.

Here we solve this model to show that introduction of such a negative
(osmotic) pressure ensures that the two wall scenario behaves in the
same way as the one wall case, in contrast to the model without any
osmotic pressure.

\section{The model}
\begin{figure}[ht!]
\begin{center}
\includegraphics[width=10cm]{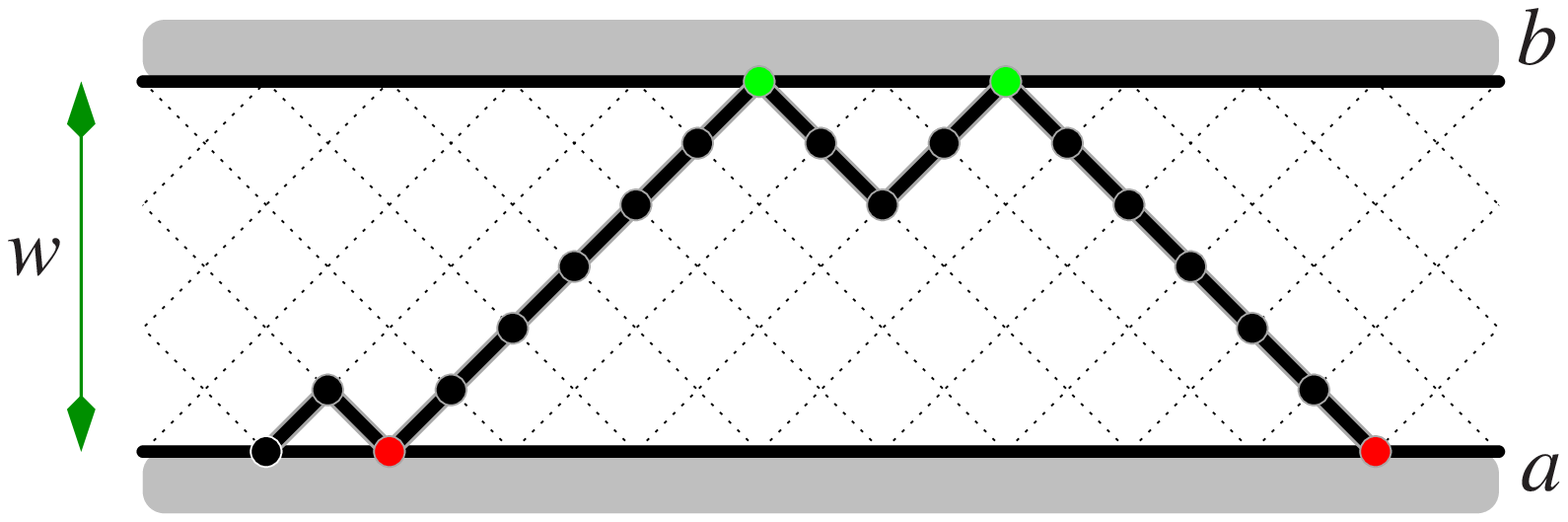}
\caption{\it  An example of a directed path which is a loop: both ends
of the walk are fixed to be on the bottom wall.}
\label{loop} 
\end{center}
\end{figure}

The directed walk model that we consider is closely related to Dyck
paths. Dyck paths are directed walks on $\mathbb{Z}^2$ starting at
$(0,0)$ and ending on the line $y=0$, which have no vertices with negative
$y$-coordinates, and which have steps in the $(1,1)$ and $(1,-1)$
directions.  We impose the additional geometrical constraint that the
paths lie in the slit of width $w$ defined by the lines $y=0$ and
$y=w$. We refer to Dyck paths that satisfy this slit constraint as
\emph{loops} (see figure \ref{loop}). 

Let $\mathcal{L}_w$ be the set of loops in the slit of width $w$. We
define the generating function of these paths as follows:
\begin{equation}
  L_w(z,a,b,p)  =   \sum_{\pi \in \mathcal{L}_w} z^{n(\pi)} a^{u(\pi)} b^{v(\pi)} p^{m(\pi)}; 
\end{equation}
where $n(\pi), u(\pi)$, $v(\pi)$ and $m(\pi)$ are the number of edges in the
path $\pi$, the number of vertices in the line $y=0$ (excluding the
zeroth vertex), the number of vertices in the line $y=w$, and
the area under the walk counted as the number of half-faces of the
lattice between the walk and the line $y=0$, respectively.

Let $\mathcal{L}_w^n$ be the
sets of loops of fixed length $n$ in the slit of
width $w$. The partition function of loops is defined as
\begin{equation}
Z_n (w;a,b,q)  =   \sum_{\pi \in \mathcal{L}_w^n} a^{u(\pi)}
b^{v(\pi)} q^{m(\pi)}\;.
\end{equation}
Hence the generating function is related to the partition functions
in the standard way, and for loops we have
\begin{equation}
  L_w(z,a,b,p)  =   \sum_{n=0}^\infty z^{n} Z_n (w;a,b,p) .
\end{equation}

We define the reduced free energy $\kappa (w;a,b,q)$ for loops for
fixed finite $w$ as
\begin{equation}
\kappa (w;a,b,p)  =  \lim_{n\rightarrow\infty} n^{-1} 
\log Z_n (w;a,b,p)\; .
\end{equation}

Consider the singularities $z_c(w;a,b,p)$ of the generating functions
$L_w(z,a,b,p)$ closest to the origin and on the positive real axis,
known as the critical points.  Given that the radii of convergence of
the generating functions are finite, which we shall demonstrate, and
since the partition functions are positive, by Pringsheim's Theorem 
(see {\it e.g.} Theorem IV.6 in \cite{Flajolet2009})
the critical points exist and are equal in value to the radii of 
convergence.  Hence the free energies exist and one can relate the 
critical points to the reduced free energies as
\begin{eqnarray}
\kappa(w;a,b,p)  =  -\log z_c(w;a,b,p).
\end{eqnarray}

\section{The half-plane}
\subsection{Preliminaries}

We first consider the case in which $w \to
\infty$, which reduces the problem to the adsorption of paths to a
wall in the half-plane with osmotic pressure. 

We begin by noting that once $w > n$ for any
finite length walk there can no longer be any visits to the top
surface so
\begin{equation}
\mathcal{L}_{w+1}^n  =  \mathcal{L}_w^n\qquad\mbox{for $w>n$.}
\end{equation}
Hence we define the sets $\mathcal{L}_{hp}^n$ as
\begin{equation}
\mathcal{L}_{hp}^n  =  \mathcal{L}_{n+1}^n \;.
\end{equation}
The limit $w \to \infty$ can therefore be taken explicitly.

Also, as a consequence of the above, for all $w > n$ we have $v(\pi)=0$
for any $\pi\in\mathcal{L}_w^n$.
Hence the partition function of loops in the half-plane can be defined as
\begin{equation}
  Z_n^{hp}(a,p)  =   \sum_{\pi \in \mathcal{L}_{hp}^n} a^{u(\pi)} p^{m(\pi)}\;.
\end{equation}

We define the generating function of loops in the half
plane via the partition function as
\begin{equation}
  L^{hp}(z,a,p)  = \sum_{n=0}^\infty z^{n} Z_n^{hp}(a,p) \;.
\end{equation}
In an analogous way to the slit we define the reduced free energy in
the half-plane for loops as
\begin{equation}
\kappa^{hp}(a,p)  =  \lim_{n\rightarrow\infty} n^{-1} 
\log Z_n^{hp}(a,p) = -\log z^{hp}_c(a,p).
\end{equation}
We note that in defining these free energies for the half-plane the
thermodynamic limit $n\rightarrow\infty$ is taken
\emph{after} the limit $w \to \infty$: we shall return to this order
of limits later.

\subsection{Exact solution for the generating function in the half-plane}

We use a standard decomposition argument to derive a functional equation for the
generating function $L^{hp}(z,a,p)$ as follows. Except
for the zero-step Dyck path with weight $1$, every Dyck path can be decomposed uniquely into a Dyck path
bracketed by a pair of up and down steps, followed by another Dyck path. The associated generating functions
are $apz^2 L^{hp}(pz,1,p)$ and $L^{hp}(z,a,p)$, respectively. This decomposition leads to the functional equation
\begin{equation}
\label{hpfuncteq}
L^{hp}(z,a,p) = 1+ apz^2 L^{hp}(pz,1,p) L^{hp}(z,a,p)\;,
\end{equation}
which one can rewrite as
\begin{equation}
\label{hpfuncteq2}
L^{hp}(z,a,p) = \frac{1}{1 -  a p z^2 L^{hp}(pz,1,p)}\;.
\end{equation}
By iterating this equation one can find a continued fraction expansion
for $L^{hp}(z,a,p)$ as
\begin{equation}
\label{hpconfrac}
L^{hp}(z,a,p)= \cfrac{1}{1- \cfrac{ a p z^2}{1-
    \cfrac{p^3z^2}{1- \cfrac{p^5z^2}{1-
        \cfrac{p^7z^2}{1-\ldots} } } } }  \;.  
\end{equation}
One can find this solution in the literature \cite{janse2000a-a}.
To find a series solution one can use a linearisation Ansatz, standard for a $q$-deformed algebraic equation such as 
Equation (\ref{hpfuncteq}). We substitute
\begin{equation}
\label{linearization}
L^{hp}(z,1,p)=\frac{H(pz,p)}{H(z,p)}
\end{equation}
into Equation (\ref{hpfuncteq}) with $a=1$, and find that $H(z,q)$ must satisfy the {\em linear} $q$-functional
equation
\begin{equation}
\label{Heqn}
H(pz,p) - H(z,p)- pz^2 H(p^2z,p)=0\;.
\end{equation}
We then solve this linear functional equation using a series in $z$,
\begin{equation}
\label{Hansatz}
H(z,p)=\sum_{n=0}^\infty c_n(p)z^n\;.
\end{equation}
This leads to the simple two-term recurrence
\begin{equation}
\label{hprecurrence}
(1-p^n)c_n - p^{2n-3} c_{n-2} =0\;.
\end{equation}
By iteration and using initial conditions $c_0 =1$ and $c_1=0$ one
finds
\begin{equation}
\label{Hsoln}
H(z,p)=\sum_{m=0}^\infty \frac{p^{2m(m-1)}}{(p^2;p^2)_m}(-pz^2)^m\;.
\end{equation}
Here
$(t;q)_n=\prod_{k=0}^{n-1}(1-tq^k)$
is the standard $q$-product. 
Using basic hypergeometric series notation \cite{Gasper2004},
we identify $H(z,p)={}_0\phi_1(-;0;p^2,-pz^2)$, where ${}_0\phi_1(-;b;q,t)$ is given by
\begin{equation}
{}_0\phi_1(-;b;q,t)=\sum_{n=0}^\infty \frac{q^{n^2-n}}{(b;q)_n(q;q)_n}t^n\;.
\end{equation}
Via (\ref{hpfuncteq2}) and (\ref{linearization}) we arrive at the final result 
\begin{align}
\label{hp-soln}
L^{hp}(z,a,p)& = \frac{H(pz,p)}{H(pz,p) -  a p z^2 H(p^2z,p) }\nonumber \\
&= \frac{H(pz,p)}{aH(z,p) +  (1-a) H(pz,p) } \;,
\end{align}
where $H(z,p)={}_0\phi_1(-;0;p^2,-pz^2)$.

\section{Exact solution for the generating functions in finite width
strips}

\begin{figure}[ht!]
\begin{center}
\includegraphics[width=10cm]{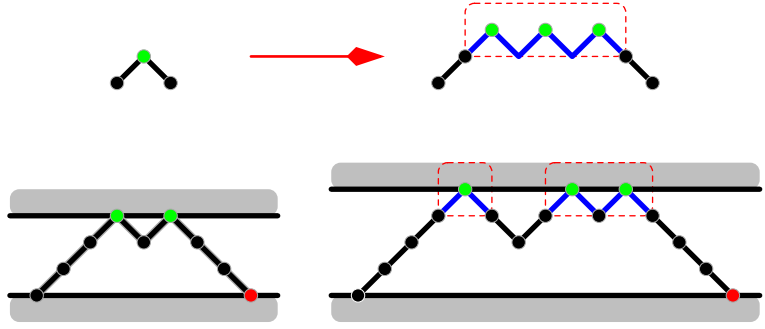}
\caption{\it  The construction of configurations in strip width $w+1$ from 
configurations in strip width $w$ is illustrated. Every vertex
touching the upper wall in the strip of width $w$ can be replaced by a 
\emph{zig-zag} path as shown.}
\label{rowbyrow} 
\end{center}
\end{figure}

The solution of the strip problem begins by using an argument that
builds up configurations (uniquely) in a strip of width $w+1$ from
configurations in a strip of width $w$. In this way a
recurrence-functional equation is constructed. Consider configurations
of loops in a strip of width $w$
(see figure \ref{rowbyrow}), and focus on the vertices touching the
top wall: call these \emph{top} vertices. These vertices contribute a
factor $b$ to the Boltzmann weight of the configuration. Now consider
a zig-zag path (see figure \ref{rowbyrow}), which is defined as a path
of any even length, or one of length zero, in a strip of width 1. The
generating function of zig-zag paths is $1/(1-bpz^2)$. Replace each of
the \emph{top} vertices in the configuration by any zig-zag path.
Since one could choose a single vertex as the zig-zag path all
configurations that fit in a strip of width $w$ are reproduced. Also,
the addition of any non-zero length path at any top vertex will result
in a new configuration of width $w+1$ and no more. The inverse process
is also well defined and so we can write recurrence-functional
equations for each of the generating functions.
The generating function $L_w(z,a,b,p)$ satisfies the following
functional recurrence:
  \begin{eqnarray}
    L_1(z,a,b,p) & = & \frac{1}{1-a b p z^2} ;\\
    L_{w}(z,a,b,p) & = & L_{w-1}\left(z,a,\frac{1}{1-bp^{2w-1}z^2},p \right).
        \label{loop-rec}  
\end{eqnarray}
We note that the zeroth vertex of the path is weighted $1$ and that
$L_w(z,a,b,p)$ counts the walk consisting of a single vertex.
We see that
\begin{equation}
 L_{2}(z,a,b,p) =\cfrac{1}{1- \cfrac{ a p z^2}{1- b p^3 z^2}}
\end{equation}
and 
\begin{equation}
 L_{3}(z,a,b,p) =\cfrac{1}{1- \cfrac{ a p z^2}{1- \cfrac{p^3 z^2}{1 -
       b p^5 z^2}}}\;.
\end{equation}
Hence, by substitution, this functional recurrence generates a finite continued
fraction expansion of the generating function
\begin{equation}
\label{confrac}
L_w(z,a,b,p)= \cfrac{1}{1- \cfrac{ a p z^2}{1-
    \cfrac{p^3z^2}{1- \cfrac{p^5z^2}{1-
        \cfrac{p^7z^2}{
          \genfrac{}{}{0pt}{}{}{\ddots 
            \genfrac{}{}{0pt}{}{}{
              \genfrac{}{}{0pt}{}{}{ 
              \genfrac{}{}{0pt}{}{}{ 
                \genfrac{}{}{0pt}{}{}{
                  \cfrac{p^{2w-3} z^2}{1-bp^{2w-1}z^2}}}}}}}}}}}\;.  
\end{equation}
This finite continued fraction can be compared directly to (\ref{hpconfrac}), which is
formally obtained by taking the limit of $w\to\infty$. Conversely, $L_w(z,a,b,p)$ is a
finitary version of $L^{hp}(z,a,p)$.

It is clear that the generating function can also be written as a
rational function
\begin{equation}
\label{Lrational1}
L_w(z,a,b,p)= \frac{P_w(z,a,b,p)}{Q_w(z,a,b,p)}
\end{equation}
though it does not simply follow to write expressions for these.
It does however follow from the theories of continued fractions and orthogonal
polynomials (see pages 256--257 of Andrews, Askey and Roy \cite{andrews1999a-a})
that both the numerator $P_w$ and denominator $Q_w$ of the generating
function satisfy recursions
\begin{equation}
\label{Precurrence}
P_w(z,a,b,p) = \begin{cases} 1\;, & w=1\\
                             1- b p^3 z^2\;, & w=2\\
                             P_{w-1}(z,a,1,p) - b p^{2w-1} z^2
                             P_{w-2}(z,a,1,p)\; & w\geq 3\;,
\end{cases}
\end{equation}
and 
\begin{equation}
\label{Qrecurrence}
Q_w(z,a,b,p) = \begin{cases} 1 - ab p z^2\;, & w=1\\
                             1- a p z^2 - bp^3 z^2\;, & w=2\\
                             Q_{w-1}(z,a,1,p) - b p^{2w-1} z^2
                             Q_{w-2}(z,a,1,p)\;, & w\geq 3\;.
\end{cases}
\end{equation}

One can immediately note that
\begin{equation}
\label{den-num-relationship}
P_w(z,a,b,p) = Q_w(z,0,b,p)
\end{equation}
so that 
\begin{equation}
\label{Lrational2}
L_w(z,a,b,p)= \frac{Q_w(z,0,b,p)}{Q_w(z,a,b,p)}\;.
\end{equation}

We now form the width generating function for the denominator as
\begin{equation}
\label{width-gen-fn}
W(t,z,a,b,p) = \sum_{w=0}^{\infty} Q_w(z,a,b,p) t^w\;,
\end{equation}
and find a functional equation for $W(t)$ from the recurrence
(\ref{Qrecurrence}) as
\begin{multline}
\label{funct-eqn-W}
W(t,z,a,b,p) =\\
 t(1-t)(1-a b p z^2)+t^2(1- a p z^2 - b p^3 z^2) 
+ t W(t,z,a,1,p) - b p^3 z^2 t^2 W(p^2 t,z,a,1,p)\;.
\end{multline}
One can first solve for $W(t,z,a,1,p)$ by iteration to give
\begin{multline}
 \label{W1-soln}
W(t,z,a,1,p) =\\
 \sum_{n=0}^{\infty} \frac{(-1)^n p^{2n^2+2n} z^n t^{2n}
\left[ t p^{2n} (1-tp^{2n}) ( 1- a p z^2) + t^2 p^{4n} (1-a p z^2- p^3 z^2)\right]}{(t;p^2)_{n+1}}
\end{multline}
It is then straightforward to provide an expression for
$W(t,z,a,b,p)$ by substituting into the functional equation
(\ref{funct-eqn-W}). To simplify the expressions further, let us 
rewrite our expression for $W(t,z,a,1,p)$ in terms of 
\begin{equation}
\label{phi}
\phi(t,q,x) 
=\sum_{n=0}^{\infty} \frac{q^{n(n-1)} x^n}{(t,q)_n}\;.
\end{equation}
Using basic hypergeometric series notation \cite{Gasper2004}, we identify
$\phi(t,q,x) ={}_1\phi_2(q;0,t;q,x)$. With help of $\phi(t,q,x)$ we can write
\begin{equation}
W(t,z,a,1,p) = \frac{1-a p z^2}{p z^2 t} - 1 
 - \frac{1-a p z^2}{p z^2 t}\phi(t,p^2,-p z^2  t^2)
+  \phi(t,p^2,-p^3z^2 t^2) 
\;,
\end{equation}
which leads to
\begin{multline}
W(t,z,a,b,p) = \frac{b + t-bt -a p z^2t - b p z^2 t -a p^2 z^4 t^3 +
  a b p z^2 t + a b p^2 z^4 t^3 -a b p z^2 }{p z^2 t} \\ 
   +\frac{(bt-b-t)(1-apz^2)}{p z^2 t}\phi(t,p^2,-pz^2 t^2)
   - (bt-b-t)\phi(t,p^2,-p^3 z^2 t^2)\; .
\end{multline}
To obtain an explicit expression for $Q_w$ one first expands the $q$-product in the function $\phi$ with the
help of the $q$-binomial theorem \cite{Gasper2004} to obtain
\begin{equation}
\label{phi-expand}
\phi(t,q,xt^2) = 1 + \sum_{m=0}^{\infty} t^m \sum_{n=1}^\infty
q^{n(n-1)}
\qbinom{m-n-1}{n-1}{q} x^n\;,
\end{equation}
where the $q$-binomial coefficient is defined as
\begin{equation}
\qbinom{N}{M}{q}=\begin{cases}\dfrac{(q;q)_N}{(q;q)_M(q;q)_{N-M}}&0\leq M\leq N\;,\\0&\text{otherwise.}\end{cases}
\end{equation}

After some algebraic manipulations this gives us an explicit and surprisingly elegant 
expression for $Q_w$ as
\begin{multline}
Q_w(z,a,b,p) =    \sum_{m=0}^\infty
(-pz^2)^m p^{2m(m-1)}\times\\
\left( (1-b) \qbinom{w-m}{m}{p^2} 
    +b \qbinom{w-m+1}{m}{p^2} 
     - (1-a)(1-b) \qbinom{w-m+1}{m-1}{p^2}
    - (1-a) b \qbinom{w-m}{m-1}{p^2} \right)\;,
\end{multline}
and hence via (\ref{Lrational2}) an explicit expression for the generation
function for loops $L_w$,
\begin{multline}
L_w(z,a,b,p) =\\
\frac
{\sum\limits_{m=0}^\infty
(-pz^2)^m p^{2m(m-1)}
\left( (1-b) \qbinom{w-m}{m}{p^2} 
    +b \qbinom{w-m+1}{m}{p^2} 
     - (1-b) \qbinom{w-m+1}{m-1}{p^2}
    - b \qbinom{w-m}{m-1}{p^2} \right)}
{\sum\limits_{m=0}^\infty
(-pz^2)^m p^{2m(m-1)}
\left( (1-b) \qbinom{w-m}{m}{p^2} 
    +b \qbinom{w-m+1}{m}{p^2} 
     - (1-a)(1-b) \qbinom{w-m+1}{m-1}{p^2}
    - (1-a) b \qbinom{w-m}{m-1}{p^2} \right)}\;.
\end{multline}
Note that the sums involved only have finitely many non-zero terms, 
as the $q$-binomial coefficients are zero when $m>\lfloor w/2+1\rfloor$.

For $a=b=1$ we obtain the particularly simple identity
\begin{equation}
L_w(z,1,1,p) =
1-\frac{\sum\limits_{m=0}^\infty(-pz^2)^m p^{2m(m-1)}\qbinom{w-m}{m-1}{p^2}}
{\sum\limits_{m=0}^\infty(-pz^2)^m p^{2m(m-1)}\qbinom{w-m+1}{m}{p^2}}\;.
\end{equation}

\section{Analysis of the free energy}
\subsection{Finite width}
When the width is finite ($w< \infty$) the system is effectively
one-dimensional and so there cannot be any phase transitions at finite
temperatures. Mathematically, one can see this in the following way. 
Regardless of whether $p=1$ or $p<1$ the generating function $L_w(z,a,b,p)$ 
is a ratio of polynomials $P_w(z,a,b,p)$ and $Q_w(z,a,b,p)$. As such the
only singularities in the generating function occur at zeros of
$Q_w(z,a,b,p)$. Given that these are orthogonal polynomials satisfying the same
three-term recurrence, albeit with different initial conditions, the zeros of 
$P_w(z,a,b,p)$ are distinct to those of $Q_w(z,a,b,p)$. The simple zero in $Q_w$ 
leads to a simple pole in $L_w$. The location of the zero $z_c(w;a,b,p)$ is 
an analytic function of $a$, $b$ and $p$.  

\subsection{Half-plane limit}
We now use that
\begin{equation}
\lim_{w \rightarrow\infty}  \qbinom{w-m}{m}{p^2} =  
\frac{1}{(p^2,p^2)_m}
\end{equation}
to give 
\begin{align}
\lim_{w \rightarrow\infty} Q_w(z,a,b,p)  &= \sum_{m=0}^{\infty}
(-pz^2)^m p^{2m(m-1)} \left[ \frac{1}{(p^2,p^2)_m} - (1-a)\frac{1}{(p^2,p^2)_{m-1}}\right]\\
&=H(z,p) -  (a-1) p z^2 H(p^2z,p)\nonumber\\
&=aH(z,p)+(1-a)H(pz,p)\nonumber
\end{align}
and
\begin{align}
\lim_{w \rightarrow\infty} P_w(z,a,b,p) &= \lim_{w \rightarrow\infty} Q_w(z,0,b,p)
= H(pz,p)\;.
\end{align}
One can therefore demonstrate explicitly that for $p<1$
\begin{equation}
\lim_{w \rightarrow\infty} L_w(z,a,b,p) = L^{hp}(a,p) =  \frac{H(pz,p)}{aH(z,p)+(1-a)H(pz,p) }\;.
\end{equation}

The series $H(z,p)$ converges absolutely for any $z$ provided $p<1$,
and so has no singularities as a function of $z$. Therefore, the only
singularity of $L^{hp}(a,p)$ occurs as a result of zeros of $H(z,p) -
(a-1) p z^2 H(p^2z,p)$ provided they don't cancel with zeros of the
numerator. As such, consider that the denominator can be written as
$H(pz,p) - a p z^2 H(p^2z,p)$ and that the numerator is $H(pz,p)$. For
small $z$ we have that $H(z,p)$ is positive since $H(0,p)=1$. Consider
the smallest positive zero of the $H(pz,p)$: let us call it $z_p$. Now
since $p<1$ we have $H(p^2z_p,p)>0$ so that the first zero of $H(pz,p)
- a p z^2 H(p^2z,p)$ must occur at some value $z_c<z_p$: that is, no
cancellation occurs.

Necessarily the closely singularity to the origin of
$L^{hp}(a,p)$ is then an analytic function of $a$. Hence there is no
phase transition as a function of $a$.

The situation for $p=1$ is different and well-known. It has been shown
\cite{brak2005a-:a} previously  that 
\begin{equation}
\lim_{w \rightarrow\infty} L_w(z,a,b,1) = L^{hp}(a,1) =
\frac{1+\sqrt{1-4z^2}}{1+\sqrt{1-4z^2} -2 a z^2}\;. 
\end{equation}
Considered as power series in $z$, the denominator (and numerator) of
the half-plane generating function only converges for $z<1/2$. For
small $a$ the algebraic singularity at $z=1/2$ is the closest to the
origin while for $a>2$ the pole arising from the zero in the
denominator is closer. This leads to a phase transition on varying
$a$.

\subsection{Infinite slit limit}

Let us consider the limit for $p<1$ of
\begin{equation}
\lim_{w \rightarrow \infty} z_c(w;a,b,p)\;.
\end{equation}

As we have discussed above $z_c(w;a,b,p)$ arises from a zero of the
polynomial $Q_w(z,a,b,p)$ while $z_c^{hp}(a,p)$ arises from a zero
of $H(z,p) - (a-1) p z^2 H(p^2z,p)$ which is analytic for all $z$.
Now we have that the limit of $Q_w(z,a,b,p)$ is the half plane
denominator $H(z,p) - (a-1) p z^2 H(p^2z,p)$ for all $z$. Hence, since
the zero at $z_c^{hp}(a,p)$ does not cancel with a zero in the
numerator as shown above we can deduce
\begin{equation}
 \lim_{w \rightarrow \infty} z_c(w;a,b,p) = z_c^{hp}(a,p)\;.
\end{equation}
 We also know that $z_c(w;a,b,p)$ is an analytic function of $a$ and
 $b$ for any fixed $w$ and $p$ and that $z_c^{hp}(a,p)$ is also
 analytic in $a$.

 Importantly, the argument fails when $p=1$ because the limit of the
 denominators for finite widths have a factor, depending on $b$, that
 \emph{does} cancel with one in the numerator.

\section{Comparison of $p<1$ with $p=1$ cases}
To summarise for $p<1$ we have just argued that the free energy of our
model is the same in the \emph{infinite slit} and the \emph{half
  plane} scenarios:
\begin{equation}
  \kappa^{inf-slit}(a,b,p) \equiv \lim_{w\rightarrow\infty} \kappa(w;a,b,p) = \kappa^{hp}(a,p)
\end{equation}
for all $a$, $b$ and that $\kappa^{hp}(a,p)$ is an analytic function
of $a$. On the other hand we know \cite{brak2005a-:a} that for
$p=1$ 
\begin{equation}
        \kappa^{inf-slit}(a,b,1) \equiv \lim_{w \rightarrow \infty} \kappa(w;a,b,1)
=\left\{ \begin{array}{ll} \log(2) & \mbox{ if } a,b
        \leq 2 \\ 
\log\left(\frac{a}{\sqrt{a-1}}\right) & \mbox{ if } a > 2 \mbox{ and }
        a>b \\ 
\log\left(\frac{b}{\sqrt{b-1}}\right) & \mbox{ otherwise.} \end{array}
\right. 
\label{inf-strip-free-energy}
\end{equation}
while that 
\begin{equation}
  \kappa^{hp}(a,1)= \left\{ \begin{array}{cc}
    \log(2) & a\leq 2 \\
    \log\left(\frac{a}{\sqrt{a-1}}\right) & a > 2 .
  \end{array} \right.
\label{halfplanefe}
\end{equation}
Hence for $b>\max(a,2)$ 
\begin{equation}
 \kappa^{inf-slit}(a,b,1) \neq \kappa^{hp}(a,1)
\end{equation}
and so the two-wall \emph{infinite slit} scenario is physically
different to the one wall \emph{half-plane} for $p=1$. 

From a physical point of view the $p=1$ case is more complicated than
the $p<1$ cases in that there are phase transitions in both the
\emph{half-plane} and \emph{infinite slit} scenarios and, crucially,
these are not all coincident. For $p<1$ the infinite slit and the
half-plane are the same. This implies that for $p<1$ (negative
pressure) in the infinite slit the top wall plays no part physically
in the behaviour of the system. For $p=1$, while the walls are a
macroscopic distance apart in the \emph{infinite slit}, the polymer can still
feel both walls, recalling that the polymer is also macroscopic in
length.

We have solved a simple model of a vesicle confined in a slit.
Mathematically the solution is interesting as it provides a new
orthogonal polynomial series associated with Dyck paths.  Physically
it demonstrates that an osmotic pressure term can remove the effect of
confinement and the effect of a contact potential with a far wall no
matter how strong the potential. Interesting further work would be to
investigate the scaling around the limit $w \rightarrow \infty $ and
$p\rightarrow 1$.

\section*{Acknowledgements} 
Financial support from the Australian Research Council via its support
for the Centre of Excellence for Mathematics and Statistics of Complex
Systems is gratefully acknowledged by the authors. A L Owczarek thanks
the School of Mathematical Sciences, Queen Mary, University of London
for hospitality.

 \end{document}